\documentclass{emulateapj}

\usepackage{apjfonts}

\newcommand{\etal}{et al.~}
\def\gsim{\lower 2pt \hbox{$\, \buildrel {\scriptstyle >}\over
{\scriptstyle \sim}\,$}}
\def\lsim{\lower 2pt \hbox{$\, \buildrel {\scriptstyle <}\over
{\scriptstyle \sim}\,$}}

\def\oviii{O~{\scriptsize VIII}}
\def\ovii{O~{\scriptsize VII}}

\def\neix{Ne~{\scriptsize IX}}
\def\nex{Ne~{\scriptsize X}}

\def\feii{Fe~{\scriptsize II}}
\def\fexvi{Fe~{\scriptsize XVI}}
\def\fexvii{Fe~{\scriptsize XVII}}
\def\fexviii{Fe~{\scriptsize XVIII}}
\def\source{4U~1820--303}

\shortauthors{Yao \etal}
\shorttitle{\uppercase{Iron abundance in the hot ISM}}

\begin{document}
\slugcomment{The Astrophysical Journal, 653, L000-L000, 2006 December 20}

\title{{\sl Chandra} Detection of \fexvii\ in Absorption: Iron Abundance in the Hot Gaseous Interstellar Medium}

\author{Yangsen Yao\altaffilmark{1}, 
        Norbert Schulz\altaffilmark{1},
        Q. Daniel Wang\altaffilmark{2},
        and Michael Nowak\altaffilmark{1}}
\altaffiltext{1}{Massachusetts Institute of Technology (MIT) Kavli Institute for Astrophysics and Space Research, 70 Vassar Street, Cambridge, MA 02139; yaoys@space.mit.edu}
\altaffiltext{2}{Department of Astronomy, University of Massachusetts, 
  Amherst, MA 01003}

\begin{abstract}

The iron depletion level and the gas-phase iron abundance in the hot 
($\sim10^6$ K) interstellar medium (ISM) are critical to our understanding 
of its energy balance as well as the thermal sputtering, 
cooling, and heating processes of dust grains. 
Here we report on the first detection 
of the \fexvii\ absorption line at 
$15.02$ \AA\ from the hot ISM in the spectrum of the low mass X-ray binary 
4U~1820--303 observed with the {\sl Chandra X-Ray Observatory}. 
By jointly analyzing 
this absorption line with those from \ovii, \oviii, 
and \neix\ ions in the same spectrum, we obtain an
abundance ratio as Fe/Ne=0.8(0.4, 2.1) in units of the Anders \& Grevesse
solar value (90\% confidence intervals).
We find that the result is robust with respect to 
different assumed gas temperature distributions. The obtained Fe/Ne abundance
ratio, albeit with large uncertainties, 
is consistent with the solar value, indicating that there is very little or no
iron depleted into dust grains, i.e., most of or all of the dust grains
have been destroyed in the hot ISM.

\end{abstract}

\section{Introduction \label{sec:intro}}

As one of the most abundant refractory metals, iron is an important  
constituent of the interstellar dust grains and is believed to have the 
greatest fraction of its atoms depleted into dust grains \citep{sof94}. 
Surveys in ultraviolet (UV) 
wavelength band indicate that $\gsim70\%$ of the iron in the cool and warm
medium of the Galactic disk/halo could be locked up in solid grains, and 
only the remains can be probed through absorption lines of gaseous 
\feii\ (see Savage \& Sembach 1996 and references
therein). Recent studies on X-ray  
absorption edges also provide evidence of iron depletion into dust grains
in the interstellar medium (ISM; Juett \etal 2006). 

Iron can be liberated from the dust to the gas phase through thermal 
sputtering. Theoretical calculations show that the sputtering caused by high 
velocity shocks ($v_s\simeq$ 50-200 ${\rm km\ s^{-1}}$) could destruct several 
$\times10\%$ of grains in the ISM, and that with the presence of a magnetic 
field, grain-grain collisions can also be a very efficient process for dust 
grain destruction, specifically in the case of low-velocity shocks 
($v_s\lsim100\ {\rm km\ s^{-1}}$; 
Draine \& Salpeter 1979; Jones \etal 1994). 
The dust grains can also be fragmented/destroyed through thermal evaporating
in the hot ISM.
The recycling of atoms back to the gas phase has been evidenced by
the difference of the elemental abundance ratios in the Galactic halo from
those in the Galactic disk, and by the ionization disparity of the shocked
material at various post-shock distance \citep{jen95, sav96}. 
A comparison between the observed
abundance ratio of (Mg+Fe)/Si in the dust of the halo clouds and that of the 
theoretical expectation, together with the observed correlation between the 
gas-to-dust ratio and the dust mass carried by Fe, indicates that the dust 
grain cores likely contain iron oxides and/or metallic iron. Some 
of these resilient cores can survive from the
shock destruction during the dust processing in the ISM \citep{sem96, fri03}.

In the hot ISM, the knowledge of the exact amount of iron contained in dust
grains, which is reflected in the gas-phase abundance of 
iron (e.g., Savage \& Sembach 1996), is of great importance for 
understanding many astrophysical processes. 
For an emission spectrum of a solar abundance plasma at 
temperature $\sim10^6$ K, the contributions of iron ions
encompass about 50\% of the total emitting energy and photons
in the energy range from 10 eV to 2 keV. 
On the other hand, dust grains are an effective coolant of the hot gas.
The thermal energy of the hot gas can be transferred to dust grains via the
collisions between electrons and dust grains, resulting in bulk heating 
of the dust and infrared (IR) emission. For a solar abundance of the dust 
grains, the grain IR radiation is at least comparable to, and could be more 
than 10 times more efficient than, the gas X-ray emission in cooling the 
hot gas at 
temperature $\sim10^6$ K \citep{dwek92}. Consequently, the existence of the
iron-bearing dust in the hot ISM could alter the chemical composition of the
hot gas, therefore, largely change the flux and the 
spectral shape of the hot gas radiation, affect the energy balance, 
change the cooling rate thus the lifetime of the hot gas, and eventually, 
adjust the star formation rate, in the whole ISM and affect the galactic
evolution in general.

The iron depletion level in the hot gas can be measured by modeling the iron 
emission lines or the absorption lines that the gas-phase iron ions imprint on 
the background point source spectrum, then comparing the iron abundance
with the solar values. Directly measuring the IR emission of the 
dust in such environments is difficult due to the confusion with  
foreground cold dust in the Galactic disk along the line of sight. 
A recent study of the very soft (0.25 keV) X-ray diffuse background emission 
in the vicinity of the Sun suggests that the gas phase iron is $\sim30\%$ 
of the solar value \citep{san01}. However, given that the Sun may reside in 
a privileged location (i.e., the Local Hot Bubble), this 
depletion level may not be typical for the general hot ISM. Furthermore,
the inferred emission measure is also sensitive to different adopted 
plasma emission models, and the less well known interaction between the solar 
wind ions with the local neutral ISM may further complicate the 
interpretations (e.g., Sanders \etal 2001;
Pepino \etal 2004; Hurwitz \etal 2005).
Absorption line studies, on the other hand, measuring the total column
density along the line of sight, are less affected by the different models and
the charge exchanges between the solar wind and the local neutral gas, and, 
therefore, should provide a reliable measurement of the iron abundance in 
the hot ISM.

In this Letter, we present the first detection of the \fexvii\ absorption line
at $\sim 15$ \AA\ from the hot ISM toward \source. 
Comparing the column density of \fexvii\ with that of \neix, and with those of 
\ovii\ and \oviii\ that we previously measured, we have derived a
relative abundance ratio of Fe/Ne and then inferred the iron depletion level
in the hot gas.
Throughout the Letter, we adopt the
solar abundances from \citet{and89}, 
\footnote{The ``real'' solar abundances are still under debate
(e.g., Asplund \etal 2005; Antia \& Basu 2006), so we still use
the old values.}
and quote the errors at 90\% confidence levels for single varying 
parameters until otherwise specified.

\begin{figure}
\centerline{
\includegraphics[width=0.47\textwidth]{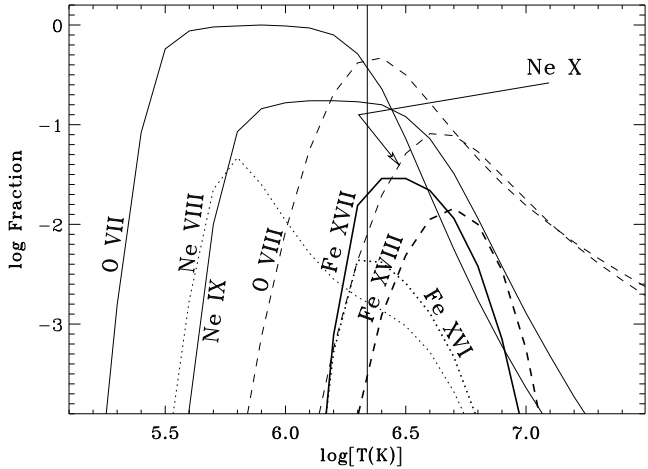}}
\caption{Ionization fractions of oxygen, neon, and iron ions as a 
   function of
   temperature for a gas in the collisional ionization equilibrium 
   state \citep{arn92}. The fractions of neon and iron have been scaled down
   with respect to the relative solar abundance ratio to oxygen.
   The {\sl vertical line} indicates the best fit temperature 
   ($2.2\times10^6$ K) in the isothermal case 
   for the absorbing gas toward \source.
\label{fig:CIE} }
\end{figure}

\section{Source, observations, and the existing results \label{sec:source}}

\source\ is a bright low mass X-ray binary residing in the globular cluster
NGC 6624 (Galactic coordinates $l, b = 2\fdg79, 7\fdg91$) and its
distance has been determined as $7.6\pm0.4$ kpc \citep{kuu03},
meaning that it is very close to the Galactic center and is located 
$\sim 1$ kpc above the disk plane.

The {\sl Chandra X-Ray Observatory} has observed this source three times
with different instrumental configurations. 
The observation log, data reduction and analysis procedures
have been described in detail in Yao \& Wang (2006, hereafter Paper I),
and here, we summarize absorption line detections and
relevant absorption line diagnostic results.

We have detected highly ionized \ovii,
\oviii, and \neix\ K$\alpha$, and \ovii\ K$\beta$ absorption lines, 
which are produced in the hot ISM rather than in the circumstellar
gas associated with the binary system (Paper I; see also Futamoto \etal 2004;
Juett \etal 2006).
Modeling these lines with our absorption line model, 
{\sl absline} \citep{yao05}, we have constrained dispersion velocity of the
hot gas [$v_b=255(165, 369)\ {\rm km\ s^{-1}}$], 
column densities of \ovii, \oviii, and \neix. We have also obtained the
abundance ratio of Ne/O in the hot gas, which is consistent with the solar 
value. For a gas at temperature about $10^6-10^7$ K, the population of each 
abundant iron ion contained in the gas, e.g., \fexvi, \fexvii, and \fexviii, 
is distributed in a narrow temperature range (Fig.~\ref{fig:CIE}); therefore,
a well confined gas temperature or its distribution is crucial for inferring 
the total iron in the hot gas. The detection of multiple absorption lines
in this sight line enables us to obtain such a constraint.
For instance, if the intervening gas is in the collisional 
ionization equilibrium (CIE) state \citep{arn92} and isothermal, 
its temperature can be determined as $T=2.2\pm0.3\times10^6$ K (Paper I).

\section{\fexvii\ absorption line and iron abundance in the hot ISM}

\begin{figure}[hb!]
\centerline{
\includegraphics[width=0.47\textwidth]{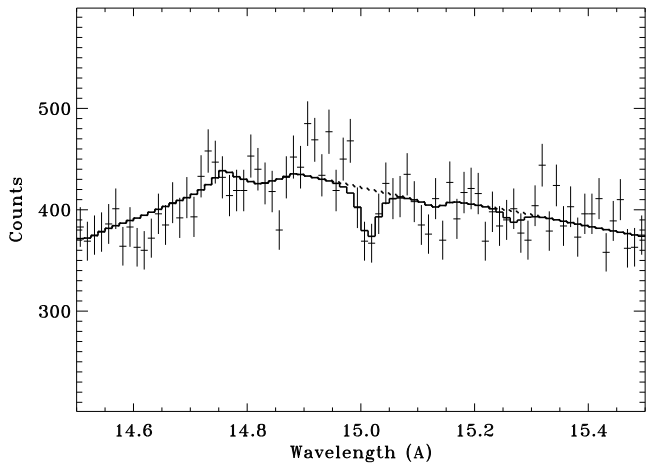}}
\caption{\fexvii\ absorption line at 15.02\AA\ in the {\sl Chandra}
   MEG-LEG combined spectrum of \source, 
   modeled with the {\sl absline} model ({\sl histogram}). 
   The {\sl dip} at 15.26 \AA\ indicates the expected amount of absorptions
   for the second most significant \fexvii\ line (Table~\ref{tab:lines}). 
   The dotted line indicates the smooth continuum level around the line. 
   The model has
   been convolved with the instrumental response, and the binsize is 
   12.5 m\AA. The apparent bump on the blue-side of the line may be due to
   the imperfect calibration near the hot pixel. For the detail information
   of the co-added spectrum and the continuum fit, please refer to Paper I.
   \label{fig:counts} }
\end{figure}

We searched for the ionized iron absorption lines at the corresponding rest 
frame wavelengths in the wavelength range between 2 and 25\AA\
in the spectrum obtained
in Paper I. The \fexvi\ absorption lines are expected to be very
weak (the oscillation strength $f_{\rm ij} < 10^{-6}$), and are not 
considered further in this work. Table~\ref{tab:lines} lists the 
strong lines ($f_{\rm ij} > 0.5$) of ions \fexvii\ and \fexviii. We only
detected a significant \fexvii\ absorption line at 15.02\AA\ 
(Fig.~\ref{fig:counts}), and did not see any clear sign for the other lines 
listed in Table~\ref{tab:lines}. These detection results are not surprising. 
The constrained hot gas temperature favors the \fexvii\ population, and 
the transition of the 15.02 \AA\ line is strongest 
(Fig.~\ref{fig:CIE}; Table~\ref{tab:lines}). Therefore,
the \fexvii\ absorption line at $\sim15.02$ \AA\ is expected to be at least 
4 times stronger (in terms of equivalent width [EW]) than the others.

\begin{deluxetable}{llcc}
\tablewidth{0pt}
\tablecaption{Strong Transition Lines of \fexvii\ and \fexviii
\label{tab:lines}}
\tablehead{
 Ion & Transition & $\lambda$ (\AA)& $f_{\rm ij}$ }
\startdata
\fexvii\ & $2s^22p^6(^1S)-2s^22p^54d(^1P^0)$ & 12.123 & 0.53 \\
\fexvii\ & $2s^22p^6(^1S)-2s^22p^53d(^1P^0)$ & 15.015 & 2.31 \\
\fexvii\ & $2s^22p^6(^1S)-2s^22p^53d(^3D^0)$ & 15.262 & 0.63 \\
\fexviii\ & $2s^22p^5(^2P^0)-2s^22p^4(^1S)3d(^2D)$ & 14.121 & 0.90 \\
\fexviii\ & $2s^22p^5(^2P^0)-2s^22p^4(^1D)3d(^2P)$ & 14.203 & 0.57 \\
\fexviii\ & $2s^22p^5(^2P^0)-2s^22p^43d(^2D)$ & 14.361 & 0.93 \\
\hline                   
\enddata                 
\end{deluxetable}

We use different models to characterize the \fexvii\ absorption 
line at 15.02\AA. The negative Gaussian model 
gives the line centroid 
as 15.008(14.999, 15.018) \AA\ or 132(-50, 314) ${\rm km\ s^{-1}}$, 
line width as
$\sigma<280\ {\rm km\ s^{-1}}$ or $v_b<396\ {\rm km\ s^{-1}}$, and its 
EW as 5.1(2.9, 7.3) mA. 

We then fit the line with our {\sl absline} model, which adopts the 
Voigt function as line profile and allows a joint analysis 
of multiple absorption lines (see Yao \& Wang 2005; Paper I; 
Wang \etal 2005 for further discussion). The fit is as good as with
a Gaussian model, and the obtained line 
position is also identical. Connecting the $v_b$
with those of \ovii, \oviii, and \neix\ lines (since the non-thermal 
broadening dominates, we therefore ignore the tiny differences of the thermal
broadening in different elements), we obtain the column density of the
\fexvii\ as $\log[N_{\rm FeXVII}({\rm cm^{-2}})]=15.0(14.7, 15.2)$.
\footnote{
This column density, together with the dispersion velocity $v_b$ constrained in
jointly analyzing oxygen and neon lines (\S2),
can reproduce the EW measured with Gaussian model.}

\begin{figure}
\centerline{
\includegraphics[width=0.47\textwidth]{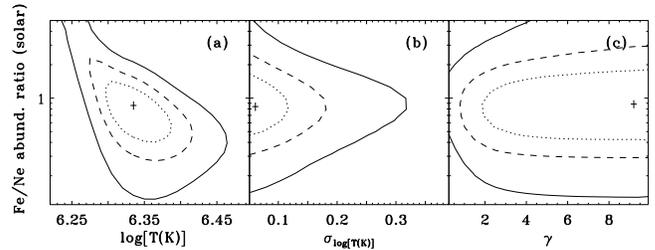}}
\caption{The 68\%, 90\%, and 99\% confidence contours of (a) the abundance 
   ratio of Fe/Ne (in units of solar value) vs. gas temperature $T$, 
   (b) Fe/Ne vs. temperature 
   dispersion, and (c) Fe/Ne vs. power-law index $\gamma$ for the 
   isothermal case, log-Gaussian temperature distribution, and the 
   power-law temperature distribution, respectively.
\label{fig:FeNe} }
\end{figure}

Following the procedure we established in Paper I, we probe the abundance ratio
of Fe/Ne in the hot gas. Since neon is a noble element, it is very unlikely 
depleted into dust grains. Therefore we take it as the reference element. 
In fact, we have obtained a Ne/O ratio that is consistent with the solar 
value (Paper I), and the following inferred Fe/Ne is essentially the same as 
Fe/O in units of solar value. Assuming that the absorbing gas is in a CIE state
and isothermal, we jointly analyze the \fexvii\ line with the \ovii, \oviii,
and \neix\ K$\alpha$ lines and the \ovii\ K$\beta$ line, requiring the common
absorbing gas to be of the same temperature. We fix the neon abundance at
the solar value, and let the abundances of oxygen and iron be free parameters.
In this way we constrain the hydrogen column density  to 
$N_{\rm H}=7.9(5.0, 10.2)\times10^{19}\ {\rm cm^{-2}}$, 
the temperature to $T=2.2(1.9, 2.5)\times10^6$ K, 
and the abundance ratio to O/Ne = 0.9(0.5, 1.3) solar
for the absorbing hot gas, which are identical to those reported in Paper I. 
In addition, we obtain the abundance ratio of Fe/Ne as 0.8(0.4, 2.1) 
solar. 
Considering the dependence of the \fexvii\ population
on $T$ (Fig.~\ref{fig:CIE}), we calculate the confidence contours of Fe/Ne 
versus $T$, which is presented in Figure~\ref{fig:FeNe}a. 


Next, we investigate the effects on the inferred Fe/Ne ratio if the above
isothermal assumption of the intervening gas is relaxed.
Since the absorption samples almost the entire Galactic disk
from the Sun into the Galactic bulge, it is possible that the hot gas 
consists of different temperature components. Here we examine two 
simple temperature distributions. In each case, we first interpolate the 
ionization fractions at different temperatures, assuming the gas to be 
in CIE state, 
and then calculate the column density for each ion.
To get a better constraint, we also add the undetected K$\alpha$ 
\nex\ (12.134 \AA) absorption line in the joint fit. 
This line, except for the detected line absorptions from oxygen, neon, and
iron ions, is the next most expected one from a different ion to be observed
in the spectrum with high counting statistic because of its anticipated large
column density (Fig.~\ref{fig:CIE}) and large transition coefficient 
($f_{\rm ij}=0.416$), and is particular useful for constraining the upper 
boundary to the gas temperature.
In the first case, we assume that the hot gas temperature distribution
follows a logarithmic Gaussian form, as a natural extension of the 
isothermal single temperature case, 
\begin{equation}
{\rm d}N_{\rm H}(T) \propto \exp\left[{\frac{-({\rm~log}T - {\rm~log}T_0)^2}{2(\sigma_{{\rm log}T})^2}}\right] {\rm d}{\rm log}(T), 
\end{equation}
where the mean temperature 
$T_{\rm 0}$ is equivalent to $T$ in the isothermal case, and 
$\sigma_{\log T}$ is the dispersion of $\log T_{\rm 0}$. Under this assumption,
we obtain $T_{\rm 0}=2.0(1.8, 2.4)\times10^6$ K and
$\sigma_{\log T}< 0.15$, and 
the abundance ratio Fe/Ne = 0.8(0.4, 2.3) solar. 
In Paper I, we have demonstrated that since an isothermal absorbing 
plasma is adequate to describe the observation, the additional free parameter
$\sigma_{\log T}$ cannot be fully constrained in the spectral fitting, 
and there is an apparent 
correlation between $T_{\rm 0}$ and $\sigma_{\log T}$ (see Fig.~4 in Paper I)
due to the large value of $N_{\rm OVIII}$. But the inferred Fe/Ne is
insensitive to different $\sigma_{\log T}$ values, as illustrated
in the confidence contours of the Fe/Ne ratio versus $\sigma_{\log T}$ 
(see Fig.~\ref{fig:FeNe}b). 

In the second case, we  assume that the hot gas temperature
distribution follows a power-law (PL) form, 
\begin{equation} \label{equ:PL}
    dN_{\rm H}(T) = \frac{N_{\rm H0}(\gamma+1)}{T_{\rm 0}} (T/T_{\rm 0})^{\gamma} dT.
\end{equation}
This simple characterization of the temperature distribution
can be derived, for example, naturally from an exponential disk model 
\citep{yao06b}, where
$N_{\rm H0}$ is the total hydrogen column density along the sight line, 
$T_{\rm 0}$ is the Galactic mid-plane temperature, and the PL index 
$\gamma$ is the ratio of the temperature to density 
scale heights. Our joint analysis gives 
$N_{\rm H}=7.5(5.3, 10.0)\times10^{19}\ \rm{cm^{-2}}$, 
$\gamma>2$, and $T_{\rm 0} = 2.4(2.1, 3.4)\times10^6$ K.
Again, although the extra free parameter $\gamma$ can vary
in a large range, the constrained abundance ratio 
Fe/Ne=0.9(0.4, 2.0) solar 
is insensitive to $\gamma$.
Fig.~\ref{fig:FeNe}c shows the confidence 
contours of the Fe/Ne ratio versus the PL index $\gamma$.

\section{Discussion}

We detect a significant \fexvii\ absorption line at $\sim15.02$ \AA\ 
in the {\sl Chandra} spectrum of 4U~1820--303. A
joint analysis of this line with the detected highly ionized oxygen and 
neon lines, all interstellar in origin and observed in
the same spectrum, gives the abundance ratio of Fe/Ne in the hot
ISM component,  which, although with large errors,
is consistent with the solar value. In addition, this result appears to be 
unaffected by the different gas temperature distributions adopted. 
We conclude that there is no evidence for substantial depletion of iron 
into dust grains in the hot ISM. This is in 
contrast to cooler phases of the ISM where iron is usually found to be
heavily depleted \citep{sem96, jue06}.
Grain cores containing iron oxides are generally rather resilient and it 
is quite difficult to liberate the iron from these cores \citep{sem96, fri03}. 
This solar value of Fe/Ne ratio, if confirmed, thus suggests that likely all 
of the dust in this
very hot ISM phase (T $\gsim 10^6$ K) has been destroyed by frequent 
and/or severe shocks during the dust grain processing in the ISM. 

Dust grains pre-existing in the ISM or formed in supernova ejecta can 
be destroyed by their generated forward and reverse shocks and 
subsequently, in heated hot gas.
By studying the iron abundance via the far-UV \feii\ absorption 
lines, \citet{sem96} find that while more than $99\%$ of iron is 
depleted into dust grains in the cold disk of the Galaxy,
the iron depletion is $\sim 80\%$ in warm clouds of the Galactic halo.
They attribute this difference to the dust grain 
disruption by the supernova (SN) shocks when circulating the grains between the 
Galactic disk and halo. If the hot ISM is believed to be heated from 
the cool ISM, it should have experienced shocks much more frequently 
and/or much more violently than the warm halo clouds. 
Therefore it is natural to expect that many more grains, even including
the resilient iron-rich cores, could have been destroyed in such harsh
environments. More recently, 
\citet{str04} have obtained an abundance ratio of Fe/O from
the diffuse extraplanar halo emission of many disk galaxies like our own,
which is $\sim40\%$ solar.
This result, although subject to different thermal plasma models adopted 
and the different thermal properties assumed for the emitting hot gas
in the data analysis, clearly rules out the depletion pattern found 
by \citet{sem96} in the cold and warm gas of the Galactic disk and halo,
further supporting the scenario that more iron has been released back
to gas phase.

The above interpretation may not be entirely unique and in some cases is 
subject to systematic
effects introduced by line-of-sight variations and intrinsic X-ray source 
properties. Claims like an overabundance of heavier elements relative
to oxygen in neutral matter towards the Galactic center direction  
presumably caused by a significant contribution of Type Ia SNe in the 
Galactic bulge \citep{ueda05} would elevate the contribution of iron in 
the hot gas phase as well.
In this respect the result by \citet{jue06} that iron is significantly 
depleted in the cool phase in line of sight towards 4U~1820--303 and that Ne/O 
is close to the solar value is an important indicator 
that conditions are not so unusual.
Abnormal heavy element abundances have been observed in microquasars 
(e.g., Cyg~X-1: Schulz \etal 2001, Juett \etal 2004; 
GRS~1915+105: Lee \etal 2002, GX 339-4: Miller \etal 2004). 
In all these systems, the observed 
overabundance pattern of the heavy elements relative to 
oxygen can also be interpreted as the different photoionization structures 
of these elements in the intrinsic absorbing material \citep{sch02, lee02}.
\citet{fut04} and Juett \etal (2004, 2006) already argued 
successfully that this is likely not the case for \source.

As a final note, we point out that the absorption path length toward
\source\ passes through the Galactic bulge region where the soft X-ray
background emission in the 0.75 keV band is greatly enhanced 
\citep{sno97}.  This enhancement indicates
that the emitting plasma is either of a high temperature or of a 
dense emitting region, or both. Recently, we have obtained an exponential 
scale height of $\sim2$ kpc and the mid-plane density of 
$\sim2.4\times10^{-3}\ {\rm cm^{-3}}$ for the hot gas toward Mrk~421 
(Galactic coordinates $l,b=179\fdg83, 65\fdg03$) in a joint-analysis 
of absorption and emission data at the same time \citep{yao06b}.
This characterization, if typical for the general hot ISM, only accounts 
for $\sim30\%$ of the observed 
absorption toward \source, meaning that a large portion of the absorption
originates from the Galactic bulge.
Therefore, the gas phase abundance ratio of Fe/Ne presented in this
letter could be biased because of the remarkable absorption contribution
from the bulge, where the metal abundance pattern may not be as the 
same as that in the overall ISM. 
Nevertheless, we present here a feasible way to infer the gas phase iron
abundance in the hot ISM that potentially affects our understanding of the 
cooling/heating process in the ISM in general. To obtain a global  
picture of the gas phase iron abundance in the hot gas, high quality 
absorption data along other sight lines that are away from the Galactic 
bulge region are therefore required.

\acknowledgments
We thank the referee for valuable suggestions that helped to improve
our presentation. We are also grateful to Claude Canizares and 
Aigen Li for useful discussions. 
This work is supported by NASA through the 
Smithsonian Astrophysical Observatory (SAO) contract SV3-73016 to
MIT for support of the {\sl Chandra} X-Ray Center, which is operated by the 
SAO for and on behalf of NASA under contract NAS 08-03060. Support from a 
{\sl Chandra} archival research grant AR6-7023X is also acknowledged.

\end{document}